# On the metallic behaviour in dilute two-dimensional systems

A.R. Hamilton[1,2], M.Y. Simmons[1,2], M. Pepper[1], and D.A. Ritchie[1]

[1] *Cavendish Laboratory, University of Cambridge, Madingley Road, Cambridge CB3 OHE, U.K.*
[2] *Semiconductor Nanofabrication Facility, School of Physics, The University of New South Wales, Sydney 2052, Australia*

(February 22, 2000)

We have studied the metallic behaviour in low density GaAs hole systems, and Si electron systems, close to the apparent two-dimensional metal–insulator transition. Two observations suggest a semi-classical origin for the metallic-like behaviour. Firstly the strength of the metallic behaviour is almost independent of the symmetry of the confining potential, and is predominantly determined by the low temperature resistivity (*i.e.* by $k_F l$). Secondly the fractional decrease in conductivity with increasing temperature depends only on $T/T_F$, independent of the carrier density, in both systems.

The scaling theory of localisation argues that there should be no metallic state in two dimensions (2D) for non-interacting electrons (or holes) [1]. This prediction has been revisited in recent years, with a growing interest in the possibility of a two dimensional metal. Experimental studies of a wide variety of low disorder 2D semiconductor systems have revealed an unexpectedly large decrease in the resistance as the temperature is lowered from $T \sim 1$ K, suggesting the existence of a 2D metal [2–6]. Although numerous theories have been put forward to explain this effect, the origins of this metallic behaviour, and the question of whether it persists to $T = 0$, are still subjects of great debate.

Recent experimental studies have shown that at high carrier densities, where interactions are weak ($r_s$ is small) and $k_F l$ is large, a transition back to insulating behaviour occurs [7–9]. These results are in agreement with the early studies of weakly interacting 2D systems which revealed a logarithmic increase in the resistance as $T \rightarrow 0$ [1,10,11]. Furthermore it has also been shown that even in the "metallic" regime in 2D hole systems, where the resistance drops sharply with decreasing temperature, quantum corrections to the conductivity are still present [12,13]. Localising corrections, due *both* to weak carrier-carrier interaction effects, *and* to phase coherent weak localisation, indicate that the system is still a Fermi liquid, and that the drop in resistance with decreasing $T$ is not quantum in origin [12]. These recent studies therefore argue against the existence of a true metallic phase even when interactions are strong ($r_s = 10$). In this paper we address the final remaining question – what then is the origin of the widely observed metallic-like drop in resistance in 2D systems?

A number of semi-classical mechanisms have been been put forward to explain this effect, including temperature dependent carrier-carrier scattering in a two-band system [14,15], and temperature dependent screening [16]. Using dilute 2D systems we investigate the relevance of these models, at low carrier densities, close to the transition from metallic-like to strongly localising behaviour.

The 2D hole system used here is formed in a high mobility modulation doped GaAs-AlGaAs heterojunction. [17]. Two samples were grown *simultaneously* on different (311)$A$ GaAs substrates: Sample A was grown on an undoped substrate, whereas sample B had an *in-situ* n+ back gate buried 1.2 $\mu$m below the hole gas. The simultaneous growth ensures that the two samples have identical layer thicknesses, doping densities and impurity densities. Measurements were performed using standard four terminal low frequency ac lockin techniques. Data were taken both for increasing and decreasing temperature, to ensure reproducibility of the results. The hole density could be varied from $0 - 10 \times 10^{10}$ cm$^{-2}$, with the metal–insulator transition occurring at a density of $\sim 1 \times 10^{10}$ cm$^{-2}$ ($r_s \sim 25$ if the hole effective mass is taken as $0.3 m_e$, or $r_s \sim 31$ if $m^* = 0.38 m_e$ [18]). The heterostructures are of exceptionally high quality, with a peak mobility of $1.1 \times 10^6$ cm$^2$V$^{-1}$s$^{-1}$.

In the first section of this work we consider the relevance of the two-band model to the metallic behaviour observed in dilute 2D systems. Figure 1(a) shows the characteristic drop in the zero magnetic field ($B$=0) resistivity $\rho$ of a 2D hole system with decreasing temperature. At this low hole density of $2 \times 10^{10}$ cm$^{-2}$ (close to, but on the metallic side of, the "metal"-insulator transition) the resistivity decreases by 50% as the temperature is reduced from 700 mK to 30 mK, with no signs of saturation as $T \rightarrow 0$.

A lack of inversion symmetry in a 2D system can lift the "spin" degeneracy of the electrons (or holes), creating two distinct bands. If the carriers in the two bands have different transport properties this can give rise to a metallic $\rho(T)$ over a limited temperature range. Indeed studies of the metallic behaviour in high density hole gas samples, far from the metal-insulator transition, show a strongly temperature dependent positive magnetoresistance ($B \perp$) that is characteristic of conduction in a two band system [15,19]. Further evidence for carrier carrier scattering causing the metallic behaviour comes from a comparison of the magnitudes of the increase of the sample resistivity with temperature, and the increase with magnetic field. In these studies it was shown that, ignoring Landau quantisation, $\rho(T = \infty) - \rho(T = 0) = \rho(B = \infty) - \rho(B = 0)$. All of these observations are consistent with the metallic behaviour resulting from scattering in a 2-band system. However these experiments have been conducted at high carrier densities, $(2 - 4) \times 10^{11} \text{cm}^{-2}$, far from the transition to strong localisation, and it is not clear that the results can be related to the behaviour observed at much lower densities where the metallic behaviour is strongest.

In contrast we investigate the metallic-like behaviour immediately in the vicinity of the transition. Here the carrier densities are an order of magnitude smaller than used in Refs. [15,19], and the interactions are strong: $r_s \simeq 17$ [18]. This low density has the additional advantage that complications due to the anisotropy and non-parabolicity of the valence band-structure which occur at higher energies are avoided. Fig. 1(b) shows the low field magnetoresistance for five different temperatures between 25 mK and 600 mK at $p_s = 2 \times 10^{10}$ cm$^{-2}$. Even though there is a factor of two drop in the zero field resistivity as $T$ decreases from 700 mK to 30 mK, the absence of a positive magnetoresistance shows that the $B = 0$ "spin-splitting" due to inversion asymmetry is extremely small. These results suggest that the two-band model of temperature dependent scattering is not appropriate at low densities, and cannot be solely responsible for describing the 2D metallic behaviour.

We now consider the effect changing the symmetry has on the metallic behaviour close to the transition. The symmetry of the potential confining the holes is determined by the electric field applied across the heterojunction. The combination of front and back-gates makes it possible to tune the symmetry of the confining potential and the strength of the inter-particle interactions independently. It is therefore possible to separate these two effects and determine which is responsible for the metallic behaviour observed in low density 2D systems. Sample A has been grown without a back-gate (i.e. $V_{bg} = 0$) and represents the least asymmetric case, i.e where the electric field across the sample is minimal. The asymmetry is increased in sample B by the incorporation of an $n+$ back-gate, which creates a large electric field across the hole gas. This built in p-i-n structure gives rise to an effective back-gate bias equivalent to the low temperature GaAs band-gap of 1.5 V [20]. The electric field, and hence asymmetry, in this sample can be further tuned over a limited range by biasing this back-gate. The requirement that leakage currents should be below 1pA limits the bias range to 1.25 - 1.7 V. It is not therefore possible to attain the flat-band condition $V_{bg}=0$ in sample B.

To quantify the effect of the back gate bias on the metallic behaviour we fix the carrier density at $p_s = 2.1 \times 10^{10}$ cm$^{-2}$, just on the metallic side of the transition. The influence of $V_{bg}$ on the potential profile and hole wavefunction is calculated self consistently [21]. The net electric field across the hole gas is the sum of the field due to the back-gate and the field due to the accumulation of the hole gas: $E = E_{bg} + ep_s/\epsilon$. Here $\epsilon$ is the dielectric constant, and $E_{bg} = V_{bg}/d$, where $d$ is the distance from the substrate to the heterointerface. The results are plotted in Figs. 2(a-c), with the solid line showing the potential profile. There is a large change in the electric field $E_{bg}$ from 0 to 14 kV/cm as the back-gate bias is altered from 0 to 1.7V. In contrast the electric field due to the fixed carrier density is constant at only 3 kV/cm.

The effect of the changes in the symmetry of the confining potential shown in Figs. 2(a-c) on the transport properties of the hole system is now examined. Fig. 2(d) shows the magnetoresistance for different $V_{bg}$. The three traces exhibit Shubnikov-de Haas oscillations with the same periodicity, confirming that the density is constant. The oscillations show only a single frequency, with no sign of the beating that occurs when two bands are occupied. In contrast to high density 2D hole systems the splitting of the heavy hole band is negligible at these low carrier densities.

To determine the influence of the electric field on the strength of the metallic behaviour, defined by the decrease in $\rho$ with decreasing temperature, we plot the temperature dependence of the $B = 0$ resistivity in Fig. 2(e). Although metallic behaviour is observed in all three cases, it is not obvious how to quantify the strength of the metal because each trace starts from a different value of $\rho$ at $T = 30$ mK. The change in resistance with $T$ is clearly largest for the strongest electric field (top trace), where $\Delta\rho = 0.04h/e^2$ from T=30 mK to 400 mK, compared to $0.02h/e^2$ for $V_{bg}$=0. However this direct comparison is complicated by the fact that the resistivity is larger for $V_{bg}$=1.7V than for $V_{bg}$=0V. This is because increasing the electric field due to the back-gate pushes the holes closer to the heterointerface, enhancing scattering from modulation doping. Thus even though the carrier density (and hence $k_F$) are constant, the effective disorder becomes larger with increasing $V_{bg}$.

Nevertheless the results clearly show that the metallic behaviour is affected by the back-gate bias. In these measurements the carrier density $p_s$, Fermi wavevector $k_F$, and strength of interparticle interactions $r_s$ are all fixed, with only the disorder related mean free path $l$ and symmetry varying. The metallic behaviour is therefore sensitive either to the symmetry of the confining potential, or to the disorder, or both.

To distinguish between these possibilities, we now compare $\rho(T)$ traces that originate from the same point at $T$=30 mK – i.e. we fix $k_Fl$ – and look for differences in the size of $\Delta\rho$ as $T$ is increased. To fix $k_Fl$ we increase the hole density from 1.8 to 2.3×10$^{10}$ cm$^{-2}$ as the back-gate bias is increased from 0 to 1.7 V. Although the carrier density and mean free path are now no longer constant in these measurements, they only vary by 30%, whereas the total electric field across the hole gas increases fivefold from 3 to 17 kV/cm. This can be seen in the plots of the potential profiles and hole wavefunctions in Fig. 3(a-c). The corresponding temperature dependent resistivity traces for the three back-gate biases are shown in Fig. 3(d). Despite the large change in the symmetry of the confining potential, the traces are almost indistinguishable. This confirms that the strength of the metallic behaviour at low densities is not determined by the symmetry of the confining potential, and is therefore not related to the degree of band splitting. Instead the data indicate that the metallic behaviour is primarily affected by changes in the disorder.

As the data show that symmetry related phenomena are not relevant, we finally consider the effects of disorder and screening. We note that the resistivity increases approximately linearly with increasing temperature for $T < 300$ mK in Fig. 3(d). This behaviour is reminiscent of temperature dependent screening in the clean limit [22]. Indeed previous experiments have shown that the metallic behaviour is consistent with $T$-dependent screening for temperatures down to 0.3 K [6]. We now re-examine this result, using different material systems and extending the measurements to lower temperatures. Fig. 4(a) shows the temperature dependence of $\rho$ at different carrier densities in the range $1.5 - 2.9 \times 10^{10}$ cm$^{-2}$ for sample B. At the lowest densities insulating behaviour is observed, with metallic-like behaviour becoming apparent as the carrier density is increased. This data is re-plotted in Fig. 4(b) to show the fractional change in conductivity $\Delta\sigma/\sigma$ against $T/T_F$. Here $\Delta\sigma$ is the change (decrease) in conductivity with increasing $T$, and $T_F$ is the Fermi temperature. All the data for the different densities collapse onto a common curve, which is linear at low $T/T_F$), and saturates at higher $T/T_F$.

The universality of this result is highlighted by a comparison with the "metallic" behaviour observed in the two-dimensional electron system of a silicon MOSFET [23]. Here we show data from a device with a peak mobility of $15 \times 10^3$ cm$^2$V$^{-1}$s$^{-1}$. the Temperature dependence of the resistivity in Figure 4(c) shows strong metallic-like behaviour, with up to a threefold decrease in $\rho$ as $T$ is reduced below $\sim 10$ K. The corresponding fractional change in conductivity is plotted in Fig. 4(d), where the data again collapse onto a common trace despite spanning an order of magnitude in carrier density. As with the 2D GaAs hole system, $\Delta\sigma/\sigma$ is linear in $T/T_F$ at low temperatures ($T/T_F < 0.2$), and flattens off at higher temperatures. The similarity to the p-GaAs data in Fig. 4(b) is remarkable. This result is even more surprising considering that the peak mobility is almost two orders of magnitude lower than for the p-GaAs, and the carrier densities are 1-2 orders of magnitude larger.

The observation of common behaviour for different charge species, in different material systems, argues against any mechanism for the metallic behaviour that is material dependent. Temperature dependent screening is therefore an appealing mechanism for the metallic behaviour as it does not depend on the details of the semiconductor system, and can cause an approximately linear decrease in conductivity with increasing temperature [22]: $\Delta\sigma/\sigma = -C(T/T_F) - D(T/T_F)^{3/2} + O(T/T_F)^2$. Furthermore, if the parameters $C$ and $D$ are density independent, the fractional change in conductivity will depend only on $T/T_F$. This would cause a collapse of the $\Delta\sigma/\sigma$ data on to a common trace, as seen in our data. However the analytical results of Ref. [22] show that $C$ and $D$ have a non-trivial density dependence, and it is not clear that this simple treatment of temperature dependent screening can fully explain the data. A more complex model recently proposed to explain the 2D metallic behaviour involves a combination of carrier freeze-out and temperature dependent screening [16], but we find no evidence for carrier freeze-out in our experiments [24].

The results presented here show that the spin-orbit related two band model of temperature dependent scattering does not describe the metallic behaviour observed in low density 2D hole gas systems. Instead we find that at low carrier densities the temperature dependence of the resistivity is predominantly determined by the low temperature resistivity (*i.e* by $k_F l$), independent of the electric field across the heterojunction. Furthermore we show that despite large differences in sample quality and 2D carrier densities, the metallic behaviour in *both* GaAs hole systems and Si electron systems takes a similar form: $\Delta\sigma/\sigma$ depends only on $T/T_F$. These observations therefore rule out interactions, the shape of the potential well and spin-orbit effects as possible origins of the metallic behaviour. Although temperature dependent screening may account for some aspects of the data, further work will be needed to fully explain the cause of the metallic behaviour in dilute 2D systems.

We thank D.E. Khmel'nitskii, D. Neilson and J.S. Thakur for many interesting discussions. This work was supported by EPSRC (U.K.).

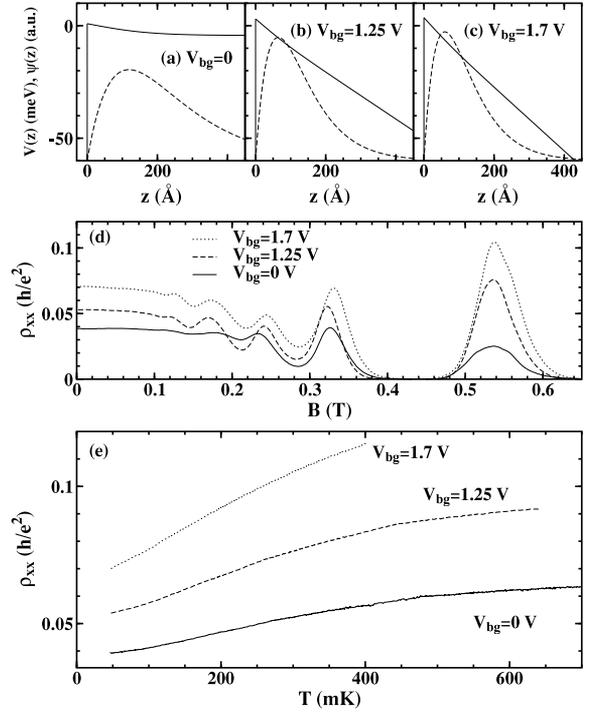

FIG. 2. Properties of the 2D hole gas at $p_s = 2.1 \times 10^{10}$ cm$^{-2}$ for different back-gate biases. (a-c) Potential profiles $V(z)$ (solid lines) and hole wavefunctions $\psi(z)$ (dashed lines). (d) Perpendicular magnetoresistance $\rho_{xx}$ measured at T=30mK. (e) Corresponding temperature dependence of the zero field resistivity.

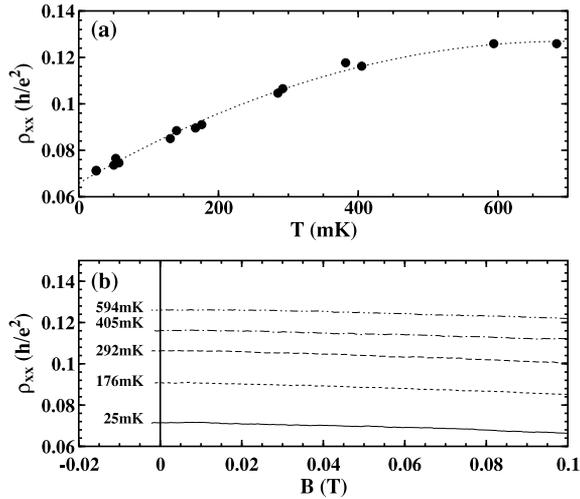

FIG. 1. (a) Temperature dependence of the zero field ($B = 0$) resistivity in the metallic regime for sample B. The carrier density is $2 \times 10^{10}$ cm$^{-2}$, and the back-gate bias is 1.7 V. (b) Low field resistivity $\rho_{xx}$ as a function of perpendicular magnetic field over the same temperature interval.

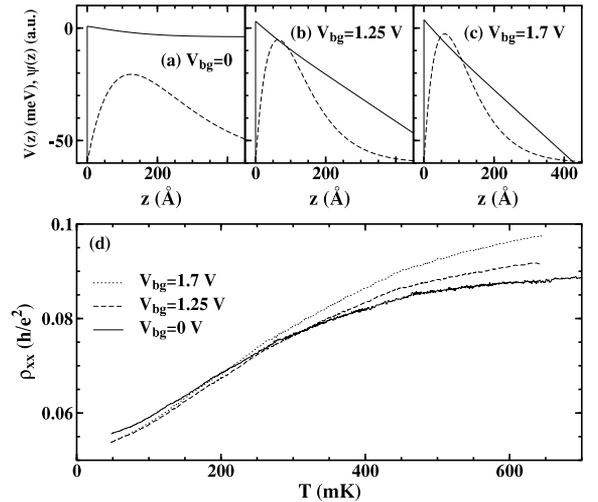

FIG. 3. Properties of the 2D hole gas with the $T = 30$ mK resistivity fixed at $0.055h/e^2$ ($k_F l = 18$), for three back-gate biases. (a-c) Potential profiles and hole wavefunctions. (d) Corresponding temperature dependence of the resistivity.

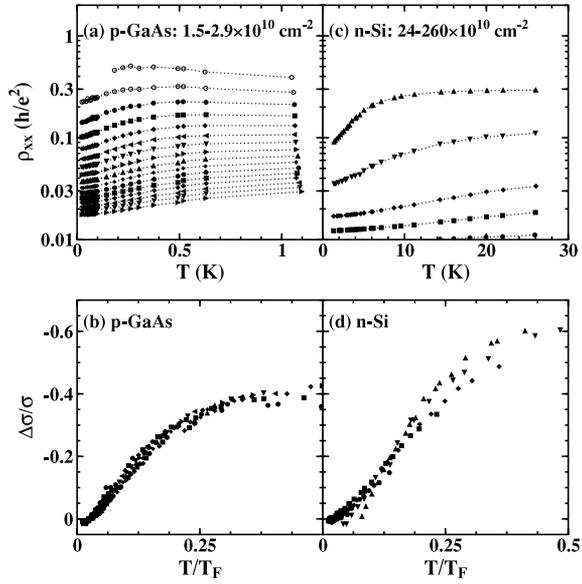

FIG. 4. Comparison of the metallic behaviour in p-GaAs and n-Si. (a) p-GaAs: Temperature dependence of the resistivity at equally spaced carrier densities in the range $1.5 - 2.9 \times 10^{10}$ cm$^{-2}$. (b) Corresponding fractional change in the conductivity $\Delta\sigma/\sigma$ as a function of the dimensionless temperature $T/T_F$. (c,d) Equivalent data for n-Si, for carrier densities of 24, 45, 100, 150 and $260 \times 10^{10}$ cm$^{-2}$